\documentclass[aps,pra,twocolumn,showpacs,amsmath,amssymb]{revtex4}
\usepackage{graphicx}% Include figure files
\usepackage{dcolumn}% Align table columns on decimal point
\usepackage{bm}% bold math
\begin{document}
                          
\title{Filamentation processes and dynamical excitation of light condensates in optical media with competing nonlinearities}
\author{David Novoa, Humberto Michinel, Daniele Tommasini and Alicia V. Carpentier}
\affiliation{Departamento de F\'{i}sica Aplicada, Facultade de Ciencias de Ourense,\\ 
Universidade de Vigo, As Lagoas s/n, Ourense, ES-32004 Spain.}

\begin{abstract}
%----------------------------   ABSTRACT  ------------------------------------
We analyze both theoretically and by means of numerical simulations the phenomena of filamentation 
and dynamical formation of self-guided nonlinear waves in media featuring competing cubic and quintic 
nonlinearities. We provide a theoretical description of recent experiments in terms of a linear stability 
analysis supported with simulations, showing the possibility of the observation of modulational 
instability suppression of intense light pulses travelling across such nonlinear media. We also show a novel 
mechanism of indirect excitation of {\em light condensates} by means of coalescence processes of nonlinear 
coherent structures produced by managed filamentation of high power laser beams.
\end{abstract}
%-----------------------------------------------------------------------------


\maketitle
%--------------------------------------------------------------------------------

\section{Introduction}
Studies on nonlinear beam filamentation of ultrashort pulses are being actively developed nowadays \cite{filamentation1}. 
Far beyond a certain critical power threshold, extremely short pulses may become unstable and break up 
into several uncorrelated light structures by modulational instability (MI) processes. These structures, 
called \emph{filaments} in this context, are spatiotemporal soliton-like light distributions which appear after MI 
and hold for large propagation distances without changing their shape and size\cite{filamentation2}. 
In fact, optical filaments have been observed in several scenarios like air\cite{filamentation1}, water\cite{dubietis03}, 
carbon disulfide\cite{centurion05}, etc., and some mechanisms to control their spatial location have been 
recently proposed\cite{Rohwetter08}. 

On the other hand, the theoretical description of these localized modes 
is in general based on nonlinear Schr\"odinger 
equations (NLSE)\cite{piekara74, dimitrievski98, Ting06}. 
In the modeling of these kind of systems, nonlinear effects like two and three photon absorption, second 
order group-velocity dispersion, time-delayed nonlinear response, multiphoton ionization and plasma 
defocusing\cite{filamentation3} are usually taken into account. Nevertheless, in order to favor the insight of dynamics 
in these systems, simpler models are employed when certain physical conditions are satisfied. To this end it is 
common to consider NLSE involving just local intensity-dependent nonlinear susceptibilities which are 
real\cite{filamentation3}. As an example, the so-called cubic-quintic (CQ) nonlinearity\cite{quiroga97} has been used 
as a solvent approximation in modelling subpicosecond pulse propagation in nonlinear media\cite{centurion05}. 
In the latter reference, the dynamics of filament formation and evolution in carbon-disulfide cells is analyzed both 
experimentally and by numerical means. At this respect, extensive numerical work based on a conservative CQ model 
is successfully employed in the qualitative description of experimental achievements. 

In addition ``liquid light condensates'', i.e, robust solitonic distributions in CQ optical media
with a ``flat-top'' transverse spatial envelope and intriguing surface tension properties were studied 
in \cite{michinel}. Recently, it was shown that their dynamics follows the same equations governing 
the evolution of usual liquid droplets\cite{novoa_PRL}. These results were obtained by both analytical and
numerical methods, but they have not yet been confirmed by real experiments. The main difficulty, of course, 
is the search for a cubic-quintic nonlinear medium, 
although some candidates were inspected in \cite{smektala00}. 
In fact, both nonlinear third and fifth order terms in the polarization of the material are usually complex, 
so it is often necessary to consider some other nonlinear processes like ones related above. 
Nevertheless, in the light of Ref.\cite{centurion05} we are allowed to think of materials like $CS_2$ as real CQ 
nonlinear media in specific power regimes. Recently, some common gases like air, $N_2$ and $O_2$ have also been 
identified as CQ optical media by means of the measurement of $n_4$ coefficients\cite{cq_demo09}, 
although the propagation dynamics of probe pulses throughout these media has not yet been investigated. 

The purpose of the present paper is threefold. Firstly, we will give a simple theoretical description 
of the results obtained in \cite{centurion05} by means of a linear stability analysis of plane waves. 
Secondly, we will study both analytically and numerically the possibility of real observation of the modulational 
instability suppression process in $CS_2$ . Finally, we will provide 
a detailed description of a novel mechanism of dynamical excitation of flat-top solitons in ``pure'' CQ systems 
via filamentation management and coalescence and we will analyze the possibility of achieving light condensates 
in carbon disulfide. 

%%%%%%%%%%%% End of Introduction %%%%%%%%%%%
\section{Physical model} 

We will consider the propagation of a high-intensity laser pulse through a $CS_2$ bulk,
 within the framework of the nonlinear Schr\"{o}dinger equation (NLSE). In order to simplify the theoretical
model, we assume a scalar slowly varying spatial envelope and we neglect group-velocity dispersion effects. 
These approximations have been experimentally justified for media like carbon 
disulfide\cite{centurion05} and air\cite{filamentation3} within proper parameter spaces. 
We will also include a cubic-quintic nonlinearity, resulting in the following NLSE

\begin{equation}\label{eq1}
2ik_{0}n_{0}\frac{\partial \psi}{\partial z}+\nabla_\perp^2 \psi+2k_{0}^2n_{0}(n_2|\psi|^2-n_4|\psi|^4)\psi=0
\end{equation}

 where the quantities above are defined as follows: $\psi$ is the complex slowly varying electric field envelope of the 
corresponding electromagnetic wave propagating through the nonlinear medium. $k_{0}=2\pi/\lambda$ is the 
vacuum wavenumber corresponding to the laser source wavelength $\lambda$. $n_{0}$ is the linear refractive index of the 
medium and $z$ is the propagation distance. Nonlinear coefficients $n_2,n_4$ characterize the strength of the real 
third and fifth order nonlinear optical susceptibilities respectively. $\nabla ^{2}_{\perp }=\partial^{2}/\partial{x^2}+
\partial^{2}/\partial{y^2}$ is the Cartesian transverse Laplacian operator. Notice that Eq.\ref{eq1} preserves the 
norm of the field $\Psi$, i.e. we have also neglected the effect of nonlinear two-photon absorption, which can be 
justified for moderate intensity pulses travelling throughout a carbon disulfide bulk over a short distance
\cite{centurion05}.
 
For technical reasons related with numerical treatment of partial differential equations, 
it is often useful to express Eq.(\ref{eq1}) in a reduced form by introducing a suitable set of non-dimensional 
variables, namely

\begin{equation}\label{eq_ad}
i\frac{\partial \Phi}{\partial \eta}+\Delta \Phi+(|\Phi|^2-|\Phi|^4)\Phi=0
\end{equation}

where the adimensional variables considered from now on are: $\eta=zk_{0}n_{2}^2/n_{4}$ is the 
adimensional propagation distance, $\Delta=\partial^{2}/\partial{\chi^2}+\partial^{2}/\partial{\zeta^2}$ 
is the dimensionless transverse Laplacian operator, being the spatial coordinates 
$\Theta=\theta k_{0}n_{2}\sqrt{2n_{0}/n_{4}}$, with $\Theta=\chi,\zeta$, $\theta=x,y$. The normalized beam 
irradiance $|\Phi|^{2}=(n_{4}/n_{2})|\psi|^{2}$ is measured in units of the quotient $n_{4}/n_{2}$ which 
univocally characterizes the nonlinear response of the medium. 

Keeping in mind the equivalence between Eqs.\ref{eq1} and \ref{eq_ad}, typical parameters corresponding to 
beam propagation in real dielectric materials like $CS_2$ are (for femtosecond optical pulses with 
$\lambda=800nm$\cite{Ganeev04}) $n_0=1.6$, $n_2=3\cdot 10^{-15}cm^2/W$. To our best knowledge, the $CS_2$
 quintic coefficient $n_4$ has not been measured yet, since there are no values of this quantity 
cited in the literature, although an approximation $n_{4}=2\cdot10^{-27}cm^4/W^2$ based on the agreement between 
simulations and experiments was successfully introduced in \cite{centurion05}. Very remarkably, the first 
experimental determination of the nonlinear coefficient $n_4$ of several gaseous media has been recently carried 
out in \cite{cq_demo09}. 

In the following section, we will introduce an analytical linear stability analysis of plane waves in the CQ 
model and its results will be compared with those of the experiments in \cite{centurion05}.

%%%%%%%%%%%% End of Physical Model %%%%%%%%%%%

%------------------------------------------------------
 
\section{Linear stability analysis of plane waves}
It is well known that Eq.\ref{eq_ad} admits plane wave (PW) 
solutions of the type $\Phi=\Phi_0e^{i\gamma \eta}$, where $\Phi_0$ is a constant which depends on $\gamma$ 
eigenvalues as follows: 

\begin{equation}
\Phi_0=\Bigg[\frac{1}{2}+\frac{\sqrt{1-4\gamma}}{2}\Bigg]^{\frac{1}{2}}
\label{A_mu}
\end{equation}

Let us proceed with a linear stability analysis of these homogeneous 
solutions, following the Bespalov-Talanov procedure\cite{bespalov66}. 
The first step of this method is to add a small 
amplitude perturbation of the form $\xi=\xi_R+i\xi_I$ to the PW solutions, 
with $\xi_{R,I}=\xi_{R,I}^0e^{[i {\bf K_{\perp}} \cdot {\bf r}+ih\eta]}$ and $|\xi_{R,I}|\ll\Phi_0$, 
where $ {\bf K_{\perp}}$ is the transverse wavevector and $h$ is the propagation constant of the corresponding 
perturbational mode. 
Now, for $h^2<0$, depending on the sign of $\Im(h)$, we have two possible 
outcomes. If $\Im(h)<0$, the perturbations grow exponentially in $\eta$ yielding to field destabilization. 
However, if $\Im(h)>0$ the amplitude of the perturbational mode decays quickly, so that the underlying 
PW remains stable. On the contrary, whenever $h^2>0$, the amplitude of the perturbations remains limited. 

%%%%%%%%%%%% figure 1 %%%%%%%%%%%%%%

\begin{figure}[tbph]
{\centering \resizebox*{\columnwidth}{!}{\includegraphics[width=0.5\textwidth]{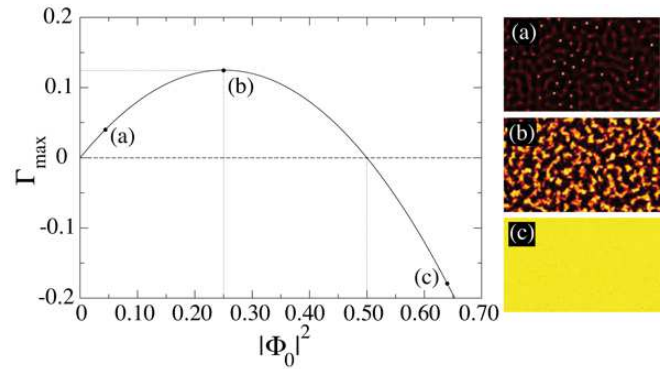}} }
\caption{ Plot of $\Gamma_{max}$ vs. $|\Phi_{0}|^2$. The highest growth rate corresponds to 
$\Phi_0^2=0.25$, whereas for $\Phi_0^2\geq0.5$ all values $\Gamma_{max}$ vanish or become negative 
delimiting the domain of modulational instability suppression. The labeled points on the graph refer to the pseudocolor 
intensity plots on the right side of the picture. These plots display a small region [width=$500$, height=$300$] 
of perturbed PW of intensities $\Phi_0^2=0.045$ (a), $\Phi_0^2=0.25$ (b), $\Phi_0^2=0.64$ (c) at the propagation 
distance $\eta=125$. The filaments shown in (b) arise before those in (a), so 
that both coalescence and domain formation effects can be observed.}
\label{fig1}
\end{figure}

%%%%%%%%%%%%%%%%%%%%%%%%%%%%%%%%%%%

By substituting the expression for the perturbed PW $\Phi=(\Phi_0+\xi)e^{i\gamma \eta}$ in Eq.\ref{eq_ad} and separating both the 
real and imaginary parts of the outgoing relation, we obtain, after linearizing on $\xi_R,\xi_I$, i.e. retaining only the first 
order terms in perturbation theory, the following couple of Schr\"odinger-type equations

\begin{subequations}
\begin{equation}\label{sch_real}
\frac{\partial \xi_I} {\partial \eta}-\Delta\xi_R-(2\Phi_0^2-4\Phi_0^4)\xi_R=0
\end{equation}

\begin{equation}\label{sch_imag}
\frac{\partial \xi_R} {\partial \eta}+\Delta\xi_I=0
\end{equation}
\end{subequations}

where $\xi_{R,I}$ depend upon $ K_{\perp}$ and $h$ in the way noted above. From Eq.\ref{sch_imag} we 
obtain $\xi_R=\frac{K_{\perp}^2}{ih}\xi_I$, and taking into account Eq.\ref{sch_real} we deduce the algebraic relation

\begin{equation}\label{Gamma}
\Gamma=K_{\perp}[-K_{\perp}^2+2\Phi_0^2-4\Phi_0^4]^{\frac{1}{2}}
\end{equation}

where $\Gamma$ is the growth rate of the perturbations, satisfying $\Gamma=ih$. Notice that 
all the information about the PW perturbational spectrum is contained in the amplitude 
$\Phi_0$. We can estimate the value of the transverse wavevector 
which gives the maximum growth rate by looking for the extremals of Eq. \ref{Gamma}, i.e

\begin{equation}\label{G}
\frac{\partial \Gamma} {\partial K_{\perp}}=0 \rightarrow (K_{\perp}^{max})^2=\Phi_0^2-2\Phi_0^4=\Gamma_{max}
\end{equation}

\begin{figure}[tbph]
{\centering \resizebox*{\columnwidth}{!}{\includegraphics[width=0.5\textwidth]{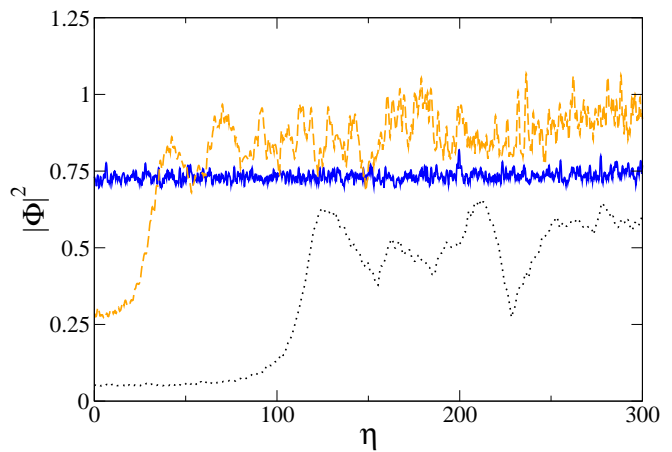}} }
\caption{ Plots of the peak intensity evolution over the line [$\zeta=0$,$-120<\chi<120$] of perturbed plane 
waves. Blue-solid line [$\Phi_0^2=0.64$] is almost constant, featuring small bounded oscillations due to spatial 
noise initially added. Orange-dashed line [$\Phi_0^2=0.25$] shows a rapid growing in intensity with $\eta$, which is a 
trace of a short longitudinal instability scale. Black-dotted line [$\Phi_0^2=0.045$] shows the destabilization scale 
for experimental data\cite{centurion05}. It is important to notice the difference in between both instability scales 
of orange-black curves. For the simulation with experimental data, the starting point of the filamentation process 
is around $\eta=75$ ($2.1mm$ in dimensional units) which is compatible with that of experiments. In all cases, the 
maximum propagation distance is $\eta=300$.}
\label{fig2}
\end{figure}

As a consequence, the PW of amplitude $\Phi_0=\frac{1}{2}$ 
features the highest perturbation growth rate (see Fig.\ref{fig1}).
This means that it will suffer destabilization 
in a shorter characteristic scale than any other homogeneous solutions of Eq. \ref{eq_ad},
if it is perturbed by a harmonic mode with $K_{\perp}=K_{\perp}^{max}$. 
These assertions are illustrated by Figs.\ref{fig1}-\ref{fig2}. In Fig.\ref{fig1} we have plotted the 
quantity $\Gamma_{max}$ as a function of the PW intensity. For $\vert\Phi_0\vert^2>0$, 
we get $\Gamma_{max}\leq0$, 
corresponding with the modulationally stable PW branch\cite{physicaD}. 
The three pseudocolor intensity plots in the right column snapshots of Fig.\ref{fig1}
display different examples 
of the state of perturbed light homogeneous distributions after a propagation distance $\eta=125$. 
All PW considered throughout this manuscript are perturbed with a $5\%$ spatial random noise and 
propagated with a standard beam propagation method. 

%%%%%%%%% figure 2 %%%%%%%%%%

%%%%%%%%%%%%%%%%%%%%%%%%%%%%%

The top picture Fig.\ref{fig1}-(a) corresponds to a simulation with $\Phi_0^2=0.045$ (unstable PW). 
It is observed that this noisy distribution leads to the formation of several self-trapped optical channels 
which survive a long distance. We have chosen this intensity because it corresponds to one of those used in the 
experiments\cite{centurion05}. We estimate the FWHM of the arising filaments to be around 
$10\mu m$, in good agreement with the experimental measurements. The middle picture Fig.\ref{fig1}-(b)
shows 
the last stage of a simulation with $\Phi_0^2=0.25$, corresponding to the PW with the highest $\Gamma_{max}$. 
In fact, as this PW posseses the shortest longitudinal scale of destabilization, it can be seen in Fig.\ref{fig1}-(b) 
that after filament formation some domains are formed due to coalescence processes between self-guiding structures. 
The final example in the bottom picture Fig.\ref{fig1}-(c) shows a stable PW of $\Phi_0^2=0.64$. 
This is a very interesting illustration of the phenomenon called {\em modulational instability suppression (MIS)}. 
For $\Phi_0^2>0.5$, the linear perturbational modes decrease exponentially, thus leading to a suppression of the 
modulational instability which could be observed in real experiments. To our knowledge, this would constitute the 
first example of beam stabilization in a nonlinear optical medium obtained just by changing the input intensity.  

Moreover, when the considered perturbations are intense enough, the quantity $\Gamma_{max}$ acquires a greater 
relevancy since its inverse value yields the characteristic longitudinal scale of instability development 
$\Lambda_{||}$\cite{bespalov66}. In a first approximation, we will consider the filamentation regime to be 
described by the coefficient $\Lambda_{||}$ as well as by the quantity $\Lambda_{\perp}=\pi/K_{\perp}^{max}$, 
which univocally characterizes the transverse spatial scale of the emerging filaments\cite{bespalov66, filamentation3}. 
We are assuming throughout this paper that optical filaments are axially symmetric so that $\Lambda_{\perp}$ is 
valid for both transverse 
coordinates ($\chi,\zeta$). In general, both quantities ($\Lambda_{||}$, $\Lambda_{\perp}$) are useful in estimating the 
order of magnitude of different spatial instability scales, but their agreement with both 
simulations and experiments can be rather qualitative due to the existence of different filamentation regimes, 
radial perturbations, nonlinear responses, inhomogeneities in the medium, coalescence processes between filaments, 
etc. In any case, as multiple filamentation appears as a consequence of modulational instability of the corresponding 
field, the linear stability analysis should be at least qualitatively correct at the leading order, where 
other nonlinear processes are not considered to be involved. 

In Fig.\ref{fig2}, we have selected a transverse cut of the three paradigmatic PW considered above 
(see Fig.\ref{fig1}) and we have determined their peak intensity along the propagation. As it can be appreciated, 
the perturbed PW with $\Phi_0^2=0.64$ (blue-solid line) features a quasi-constant intensity during 
the whole simulation, illustrating the MIS process. However, for $\Phi_0^2=0.25$ (orange-dashed) and 
$\Phi_0^2=0.045$ (black-dotted), a dramatic growing of the peak intensity with $\eta$ is observed, indicating the 
destabilization of the initial conditions. Furthermore, 
there is a substantial difference in the longitudinal instability scale of both situations. It can be appreciated 
that $\Phi_0^2=0.25$ (PW featuring the highest $\Gamma_{max}$) displays shorter instability scale than $\Phi_0^2=0.045$. 
Hereafter, we will compare this numerical result with the theory as well as the predictions of the stability analysis 
for the experimental parameters considered by Centurion {\em et al.} \cite{centurion05},
that observed the filamentation process undergone by 
femtosecond pulses propagating in $CS_2$. Using pulses of peak intensity 
$I_p=0.68$ $GW/mm^2$, they measured a mean filament sizes of $12\mu m$, in good agreement with our 
estimation.  
Moreover, the transverse scale of the filaments predicted by our linear stability 
analysis for their experimental configuration 
is $\Lambda_{\perp}=\pi/(\Phi_0^2-2\Phi_0^4)^\frac{1}{2}$, or 
$\Lambda_{\perp}\approx16.45\mu m$ in dimensional units, 
which is also consistent with the sizes measured in \cite{centurion05}. 
This can be interpreted as a further theoretical proof of the validity 
of the CQ model in describing the dynamics of femtosecond optical pulses propagating in $CS_2$. 

On the other hand, the longitudinal scale of instability development $\Lambda_{||}$ can be estimated as 
$\Lambda_{||}\approx\Gamma_{max}^{-1}=24.4$ ($0.7mm$ in dimensional units). 
This scale has not been directly characterized by Centurion {\em et al.} in Ref. \cite{centurion05}. 
Nevertheless, they 
found completely developed filaments on a scale of $5mm$, 
which is compatible with our value of $\Lambda_{||}$,
that constitutes a rough estimation of the propagation distance 
for the filaments when they start to grow up. 
Note that for $\Phi_0^2=0.25$
we get $\Lambda_{||}=8$, thus justifying the shorter characteristic scale of instability development 
observed in Fig.\ref{fig2}. 

While our analytical and numerical results for $\Lambda_{\perp}$ are both in good
agreement with the experiment of Ref. \cite{centurion05}, the
estimations for $\Lambda_{||}$ turn out to be
less accurate. This is due to the fact that in the simulations we have perturbed the homogeneous 
distributions with random noise instead of using harmonic perturbations with fixed $K_{\perp}$.
In fact, we have checked that using harmonic perturbations leads to smaller values for
$\Lambda_{||}$ by a factor of $\sim3$, getting closer to the analytical calculation.  

These results show that, in spite of its simplicity, 
the linear stability analysis not only provides 
a good insight into the physics, but it can also be used 
to obtain quantitative useful predictions for the experiments 
of filamentation in $CS_2$.

%------------------------------------------------------------------------------------

%%%%%% End of Linear stability analysis of plane waves %%%%%%

%%%%%%%%%% figure 3 %%%%%%%%%%%

\begin{figure}[tbpH]
{\centering \resizebox*{\columnwidth}{!}{\includegraphics{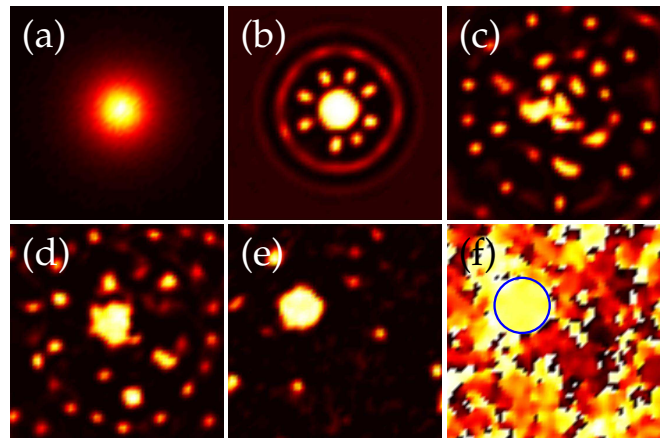}} }
\caption{ Intensity pseudocolor plots of the $SGF$ evolution in the presence of an unstable 
homogeneous background. The intensity snapshots correspond to a spatial domain [$-75<\chi,\zeta<75$] at several 
propagation distances, namely: (a) $\eta=0$; (b) $\eta=80$; (c) $\eta=150$; (d) $\eta=200$; (e) $\eta=1000$. 
Last picture (f) displays the phase map of the field depicted in (e). From this phase structure, it can be stated 
that the light condensate appearing at the last stage of the propagation is a coherent light structure since the 
color within the region where it is located (see blue circle in picture (f)) is homogeneous.}
\label{fig3}
\end{figure}

%%%%%%%%%%%%%%%%%%%%%%%%%%%%%%%

\section{Liquid light solitons arising by coalescence}

In this section we will show through numerical simulations a novel mechanism for the dynamical excitation of so-called 
``liquid light condensates''\cite{michinel} in media with CQ nonlinearity, by means of spatial control of the 
filamentation regime and coalescence processes between self-guided optical channels. Furthermore, we will discuss 
about the possibility of observation of these structures in the experiments with $CS_2$. 

As starting point, we have considered an initial condition for our simulations consisting on a broad-extended 
laser pulse with random noise fluctuations added. Group-velocity dispersion effects shall not be considered 
here since it is justified to assume that the temporal profile of the pulse does not change during short 
propagation distances through $CS_2$\cite{centurion05}. We will model this light distribution with a wide flat-top 
profile represented by the following function;

\begin{equation}
\Omega=A_{bg}\Big[0.25\Big(\Big[1+\tanh(\rho+\omega_{bg})\Big]\Big[1-\tanh(\rho-\omega_{bg})\Big]\Big)\Big]
\label{topflat}
\end{equation}

where $A_{bg}$ is the peak amplitude of the light pulse, $\rho=\sqrt{\chi^2+\zeta^2}$ is the radial coordinate 
and $\omega_{bg}$ is the mean radius. We have performed all simulations fixing $A_{bg}=0.2$ and $\omega_{bg}=1000$. 
As we will restrict ourselves to a small region located at the center of the pulse, we can think 
of the domain delimited by $\Omega$ as part of a plane wave, since boundary effects (e.g. near-field diffraction) 
will not affect our results within the longitudinal scale imposed by the common experimental setups. 
We will explore the case $A_{bg}=0.2$, corresponding to a modulationally unstable plane wave\cite{physicaD} 
as we have discussed above, although the same qualitative behavior is obtained whenever $\Gamma_{max}>0$ 
(see Fig.\ref{fig1}).

By adding an inhomogeneity to the light distribution of Eq. 
(\ref{topflat}), we will be able to manage 
the filamentation regime. In fact, we have induced a \emph{small scale Gaussian fluctuation (SGF)} of amplitude 
$A_{G}=0.5071$ and width $\omega_{G}=30$, located at the center of the bidimensional spatial profile $\Omega$ in 
order to select a small region where filamentation will first appear. Notice the scale difference 
between the two counterparts. Then by overlapping both structures $\Omega$ and $SGF$, we are generating a spatially 
inhomogeneous profile $\Phi_{in}$ with a peak intensity $|\Phi_{in}|^2=0.5$, which corresponds to that of the plane-wave 
defining the boundary between the stable and unstable PW solutions of Eq.\ref{eq1} \cite{physicaD}.

The evolution dynamics of such light distribution is illustrated in Fig.\ref{fig3}. In this figure, the snapshots show 
our small region of interest [$-75<\chi,\zeta<75$]. The initial condition is plotted in picture (a). Once the pulse 
has entered the medium, the $SGF$ starts to self-focus. It is remarkable how such a process is fastly counteracted by 
the self-defocusing nonlinearity. As a result, a bright hot spot is observed at the distribution centroid and diffractive 
rings representing shockwaves due to ``background-fluctuation'' interplay appear (see (b) snapshot). 

The spectral properties of these waves (usually referred as \emph{conical emissions}) have been 
studied in \cite{Nibbering96}. They give support to the formation of new filaments from pulse MI. 
Hereafter, we will discuss how to control the shockwave propagation in order to stimulate the appearance of 
several filaments surrounding the hot spot (see (b) snapshot). The key point is to generate a $SGF$ in a precise 
way by using phase masks\cite{Rohwetter08} and techniques like pulse shaping \cite{pulse_shaping}. 
To excite wide liquid solitons, we propose to surround the main custom-generated light structure, which is roughly 
similar to a top-flat beam, with a critical number of filaments located close of it. In fact, the solitons are 
statistical attractors due to their high internal energy. The {\em satellite filaments} are then used to provide an 
energy source for the central seed so that its power could be increased by coalescence processes, if the phases of 
the soliton and the filament match. As shown in snapshot (c) in Fig.\ref{fig3}, there is indeed a dynamical energy 
exchange between filaments and hot spot during propagation. As a result, all solitonic structures are both created 
and annihilated several times (picture (c)), until a wider coherent structure with almost constant peak intensity 
arises (picture (d)). As a consequence, the emerging liquid light soliton\cite{michinel} is strongly perturbated by the energy excess 
released in the collective coalescence. Nevertheless, from snapshot (e) we see that after a large propagation 
distance the flat-topped soliton features an improved radial symmetry, the energy excess being radiated in the form 
of linear modes.

%%%%%% figure 4 %%%%%%%

\begin{figure}[tbph]
{\centering \resizebox*{\columnwidth}{!}{\includegraphics{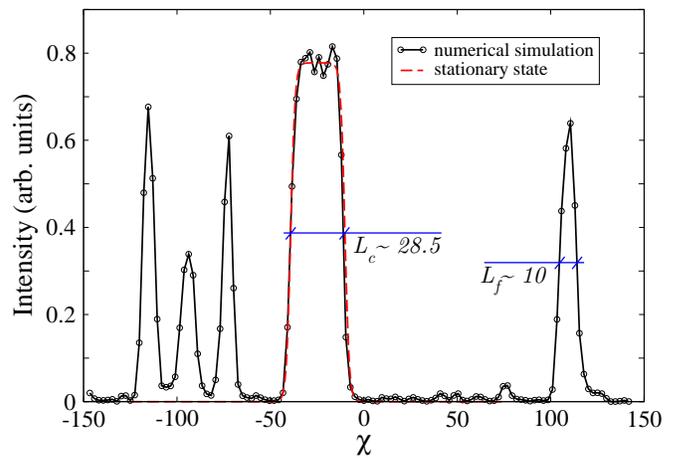}} }
\caption{ Transverse intensity profile of the filaments distribution (black solid line) at $\eta=5000$, with 
a superposed transverse profile of an eigenstate (red dashed line) of Eq. \ref{eq_ad} with propagation constant 
$\gamma=0.173$. It is seen that the agreement between both contours is reasonable. $FWHM$ (delimited with blue solid lines) of the different-sized filaments represented in the figure have been estimated to be $L_c\approx28.5$ (flat-topped soliton) 
and $L_f\approx10$ (Kerr-type soliton). The spatial coordinate $\chi\in[-150,150]$.}
\label{fig4}
\end{figure}

%%%%%%%%%%%%%%%%%%%%%%

In picture (f) we also show the phase map corresponding to the 
intensity plot displayed in (e). We see that within the region where the condensate is placed (highlighted 
with a blue circle) the phase structure is homogeneous. 

Finally,  
we have checked the 
transverse size of the distribution after a huge propagation distance, $\eta=5000$. 
This allows the light 
condensate to relax towards a stationary state, thus minimizing random profile variations due to surface modes.
In Fig.\ref{fig4}, we have compared this extracted profile with that of a stationary solution of Eq.\ref{eq_ad}. 
The agreement between the 
transverse profiles corresponding to this simulation and the stationary mode is remarkable. 
This completes the demonstration that this nonlinear structure observed in our simulations is a true flat-topped CQ-soliton.

To determine whether this kind of top-flatted transverse modes have been observed in the
experiments with $CS_2$\cite{centurion05},
we have estimated their transverse size, and found that
it would lie around the value $L_c\approx28.5$ 
($30.2\mu m$ in dimensional units). This is almost $3$ times greater than the size 
of the optical filaments measured 
in Ref. \cite{centurion05}. We thus conclude that Centurion {\em et al.} have 
not reached the threshold to produce top-flatted, liquid light solitons.

On the other hand, on the right side of the light condensate 
displayed in Fig.\ref{fig4} we also see a completely developed self-guided 
light channel of size $L_f\approx10$ ($10.6\mu m$ 
in dimensional units), which resembles closely to those observed 
by Centurion {\em et al.}. This is a further demonstration that our simulations 
reasonably reproduce the existing observations, 
besides providing a guide for the new managed production
of the higher power top-flatted solitons.

%------------------------------------------------------------------------

%%%%%% End of Liquid light solitons arising by coalescence %%%%%%
 
\section{Conclusions}

In this paper, we have provided theoretical support to recent experiments on filamentation 
of ultrashort pulses propagating through carbon disulfide by means of approximate analytical methods and numerical 
simulations of propagation. The agreement between our calculations and the experimental outcomes is remarkable, 
taking into account that we have developed the theoretical description assuming a quintic coefficient $n_4$ which 
has not been measured in the laboratory. We have shown that under certain conditions, a suppression of the modulational 
instability could be observed in such media over an intensity threshold. We have also described a novel procedure 
for indirect excitation of flat-topped solitons in pure CQ materials, and we have argued that these nonlinear 
structures have not been observed experimentally yet. Very remarkably, recent experiments with air and other 
gases have revealed their CQ nature by measuring their high-order nonlinear coefficients\cite{cq_demo09}. 
Thus our results open the door to the quest for {\em liquid light condensates} in real experiments.

%--------------------------------------------------------------------------

%%%%%% End of Conclusions %%%%%%

\section*{ACKNOWLEDGEMENTS} D.N. is very grateful to Prof. Ignacio Cirac and the whole ``Theory group'' at the 
MPQ for their warm hospitality during his stay in Garching, where part of this work was done. D.T thanks the Intertech 
group at Universidad Polit\'ecnica de Valencia for hospitality during a stay which was supported by a fellowship of 
the Universidad of Vigo. This work was supported by MEC, Spain (projects FIS2006-04190 and FIS2007-62560) and 
Xunta de Galicia (project PGIDIT04TIC383001PR). D.N. acknowledges support from Conseller\'{i}a de Innovaci\'on e 
Industria-Xunta de Galicia through the ``Maria Barbeito'' program.

%--------------------------------------------------------------------------

%%%%% End of Acknowledgements %%%%%

\end{document}